\documentclass[applemac]{aa}
\usepackage{inputenc}
\usepackage{natbib}
\usepackage{txfonts}
\usepackage{graphicx}
\usepackage{color}
\bibpunct{(}{)}{;}{a}{}{,}

\begin{document}

\title{Hydrodynamics with gas-grain chemistry and radiative transfer: comparing dynamical and static models}
\author{O. Sipil\"a
	     \and{P. Caselli}
}
\institute{Max-Planck-Institute for Extraterrestrial Physics (MPE), Giessenbachstr. 1, 85748 Garching, Germany \\
e-mail: \texttt{osipila@mpe.mpg.de}
}

\date{Received / Accepted}

\authorrunning{O. Sipil\"a et al.}

\abstract
{We study the evolution of chemical abundance gradients using dynamical and static models of starless cores.}
{We aim to quantify if the chemical abundance gradients given by a dynamical model of core collapse including time-dependent changes in density and temperature differ greatly from abundances derived from static models where the density and temperature structures of the core are kept fixed as the chemistry evolves.}
{We developed a new one-dimensional spherically symmetric hydrodynamics code that couples the hydrodynamics equations with a comprehensive time-dependent gas-grain chemical model, including deuterium and spin-state chemistry, and radiative transfer calculations to derive self-consistent time-dependent chemical abundance gradients. We apply the code to model the collapse of a starless core up to the point when the infall flow becomes supersonic.}
{The abundances predicted by the dynamical and static models are almost identical at early times during the quiescent phase of core evolution. After the onset of core collapse the results from the two models begin to diverge: at late times the static model generally underestimates abundances in the high-density regions near the core center, and overestimates them in the outer parts of the core. Deuterated species are clearly overproduced by the static model near the center of the model core. On the other hand, simulated lines of $\rm NH_3$ and $\rm N_2H^+$ are brighter in the dynamical model because they originate in the central part of the core where the dynamical model predicts higher abundances than the static model. The reason for these differences is that the static model ignores the history of the density and temperature profiles which has a large impact on the abundances, and therefore on the molecular lines. Our results also indicate that the use of a very limited chemical network in hydrodynamical simulations may lead to an overestimate of the collapse timescale, and in some cases may prevent the collapse altogether. Limiting the set of molecular coolants has a similar effect. In our model, most of the line cooling near the center of the core is due to HCN, CO, and NO.}
{Our results show that the use of a static physical model is not a reliable method of simulating chemical abundances in starless cores after the onset of gravitational collapse. The abundance differences between the dynamical and static models translate to large differences in line emission profiles, showing that the difference between the models is at the observable level. The adoption of complex chemistry and a comprehensive set of cooling molecules is necessary to model the collapse adequately.}

\keywords{astrochemistry -- hydrodynamics -- ISM: abundances -- ISM: clouds -- ISM: molecules -- radiative transfer}

\maketitle

\section{Introduction}

Models of interstellar chemistry are invoked to understand the chemical origin of line emission and absorption toward a variety of objects. Let us assume that an emission line is observed toward an object and one wants to derive the abundance distribution of the emitting molecules. The simplest approach is to assume that the molecule has a constant abundance along the line of sight, and to find the abundance that best fits the observation \citep[e.g.,][]{Jorgensen02,Tafalla04a}. Alternatively, one can assume an exponentially decaying or power-law abundance profile \citep[e.g.,][]{Tafalla02,Crapsi07}. A more elaborate approach is to use a chemical model to derive radially and temporally varying chemical abundance profiles, which is possible if the physical structure of the object is known \citep[e.g.][]{Coutens14,Harju17b}. In this case it is usually assumed that the physical structure of the object remains static as the chemistry evolves. Natural questions to ask are whether this assumption is valid and whether the simulated chemical abundances would turn out different if one took the dynamics into account. We explore these issues in the present paper. In what follows, we refer to models that keep the physical structure constant as the chemistry evolves simply as ``static models''.

Static models have trouble reproducing observations of nitrogen-bearing species, in particular ammonia. The binding energy of ammonia onto water ice is very high (canonically $\sim$5500\,K; \citealt{Collings04}), and so ammonia is expected to deplete strongly onto grain surfaces at high density and low temperature, corresponding to the centers of starless and prestellar cores. Indeed, several gas-grain chemical models show such strong ammonia depletion \citep{Semenov10,LeGal14, Sipila15b}. Observations on the other hand show that ammonia is not depleted toward the centers of starless and prestellar cores \citep[e.g.][]{Tafalla02,Crapsi07}. Because of the high binding energy, thermal or cosmic ray induced desorption are not strong enough to desorb ammonia from the grain surfaces. Another non-thermal alternative for desorption is related to the energy released in exothermic grain-surface reactions, the so-called chemical desorption process (\citealt{Garrod07}; more recently \citealt{Minissale16a,Vasyunin17}), but the inclusion of this process in a static model can produce too much gas-phase ammonia in lower-density gas away from the core center \citep{Caselli17}. The discrepancy between modeling and observations is still a mystery.

In the present work we discuss chemical abundance profiles calculated in the framework of a spherically symmetric one-dimensional hydrodynamical model and compare the results against those obtained from static models. One-dimensional hydrodynamical simulations of the collapse of starless cores in a similar context have been previously carried out by several groups, with varying degrees of complexity in terms of chemistry, for example: no chemistry \citep{Masunaga00}; limited gas-phase chemistry \citep{Keto08}; full gas-grain chemistry including deuterium chemistry \citep{Aikawa12}. Here we adopt an extensive chemical reaction scheme in the gas and on the grains that includes deuterium and spin-state chemistry, with an explicit treatment of the ortho and para states of ammonia and water \citep{Sipila15a,Sipila15b}, which is important for understanding observations (ammonia) and for an accurate treatment of line cooling (water). We combine the hydrodynamical and chemical calculations with radiative transfer methods so that the effect of chemistry on line cooling can be determined self-consistently. Our specific aim is to quantify the differences in chemical abundance profiles arising from the use of a dynamical instead of a static core model.

The paper is organized as follows. In Sect.\,\ref{s:model} we describe the details of our hydrodynamical, chemical and radiative transfer models. In Sect.\,\ref{s:results} we present the results of our modeling and discuss them in Sect.\,\ref{s:discussion}. Conclusions are drawn in Sect.\,\ref{s:conclusions}. In Appendix~\ref{appendixa} we present simple gas-phase and grain-surface chemical networks.

\section{Model description}\label{s:model}

\subsection{Hydrodynamics}

We developed a code that solves partial differential equations of hydrodynamics in one spatial dimension in the Lagrangian formalism, assuming spherical symmetry. The problem is described by the following equations:
\begin{equation}\label{eq_u}
\frac{\partial u}{\partial t} = -\frac{1}{\rho_0} \left(\frac{R}{r}\right)^2 \frac{\partial (p+q)}{\partial r} - \frac{GM}{r^2}
\end{equation}
\begin{equation}\label{eq_r}
\frac{\partial R}{\partial t} = u
\end{equation}
\begin{equation}\label{eq_rho}
\frac{1}{\rho} = \frac{1}{\rho_0} \left(\frac{R}{r}\right)^2 \frac{\partial R}{\partial r}
\end{equation}
\begin{equation}\label{eq_mass}
M(R) = 4\pi\int_0^{R} \rho R^{\prime 2}\,dR^{\prime}
\end{equation}
\begin{equation}\label{eq_E}
\frac{\partial E}{\partial t} = -(p + q) \frac{\partial}{\partial t}\left(\frac{1}{\rho}\right) + \frac{\Lambda}{\rho}
\end{equation}
\begin{equation}\label{eq_intene}
E = \frac{5}{2}R_{\rm sp}T
\end{equation}
\begin{equation}\label{eq_P}
p = \rho R_{\rm sp} T
\end{equation}
\begin{equation}\label{eq_q}
q = 
\begin{cases}
l^2 \, \rho \, (\partial u / \partial R)^2 &\partial u / \partial R < 0 \\ 
0 &\partial u / \partial R \geq 0 \, .
\end{cases}
\end{equation}
These equations represent: (\ref{eq_u}) the time evolution of the velocity field in the core ($G$ is the gravitational constant); (\ref{eq_r}) the location of a given grid cell as a function of time (in the equations, $r$ represents the original location of the cell while $R(t)$ represents its location in the moving frame); (\ref{eq_rho}) the scaling of the density profile as the core evolves ($\rho_0(r)$ represents the original density profile); (\ref{eq_mass}) the core mass as a function of radius; (\ref{eq_E}) the time evolution of the internal energy of the gas ($\Lambda$ parametrizes heating/cooling; see below); (\ref{eq_intene}) the internal energy, assuming diatomic gas ($R_{\rm sp} = k/m$ is the specific gas constant, where $k$ and $m$ are the Boltzmann constant and the average molecular mass of the gas); (\ref{eq_P}) the thermal pressure; (\ref{eq_q}) the pseudo-viscosity of the medium ($l = a\Delta R$, where $a$ is a dimensionless constant; we take $a = \sqrt3$). The above equations are essentially the same as those given in \citet{Richtmyer67}, although we have added the mass and cooling terms (see \citealt{Keto05} for a similar treatment of the problem). Equation (\ref{eq_rho}) can be substituted into (\ref{eq_u}) to consider the spatial derivative of $R$ instead of~$r$.

The workflow of our code consists of solving Eqs.\,(\ref{eq_u}) to (\ref{eq_q}) in sequence. The maximum length of the time step is set by the Courant condition for this problem:
\begin{equation}
\Delta t = C \, \frac{\Delta R}{c_s} \, ,
\end{equation}
where C is a dimensionless constant and $c_s = \sqrt{kT/m}$ is the sound speed. In practice the value of C is limited to $0 < C \lesssim 1$; for larger values the solution of the hydrodynamics equations will diverge \citep{Richtmyer67}. In this paper we set $C = 0.6$. Another factor limiting the length of the time step is the spacing $\Delta R$ of the grid cells. The choice of the initial time step, i.e., the adopted amount of grid cells, is discussed in Sect.\,\ref{ss:iso}. As the core starts to contract, the time step becomes shorter.

We use a simple first-order finite differencing method to solve the partial differential equations. The practical application of this method is detailed in \citet{Richtmyer67}, and is not repeated here.

The origin is not included in the grid considered in the calculations. We assume that the infall velocity is zero at the origin, which is reasonable given the symmetry of the problem and our first-order approach to solving the hydrodynamics equations. This assumption results in an infall profile that is consistent with previous models of the collapse of a spherical cloud \citep[e.g.,][]{Foster93, Ogino99, Aikawa05, Keto05} which show that the infall velocity is close to zero near the origin before the formation of a central object. Here we do not attempt to follow the dynamical evolution all the way to the formation of a protostar; the calculational loop is designed to break when the flow becomes supersonic. If the supersonic condition is not reached until $t = 5 \times 10^6\,\rm yr$, the code terminates.

\subsection{Integration of chemistry and radiative transfer}\label{ss:int_chem_rt}

In this work we incorporate a self-consistent treatment of gas-grain chemistry into the framework of hydrodynamics. Chemical abundances evolve in a time-dependent fashion in tandem with the physical evolution of the core. In particular, chemistry influences the dynamic evolution of the core through its effect on the line cooling. The chemical evolution is calculated in a subset of the full grid (25 cells), which are distributed in such a way that the grid resolution is higher near the origin than at the the outer, less dense parts of the core where chemical evolution is slow. Running the chemical calculations in a subset of the full grid is also imperative to keep the total run time of the model in a manageable scope (see Sect.\,\ref{ss:codeeff}).

We use the chemical model described in detail in \citet{Sipila15a} and \citet{Sipila15b}. In short, the code considers gas-grain chemistry assuming that the ice on the grain surfaces consists of a single reactive layer, and that the grains themselves are spherical with a radius of 0.1\,$\mu$m. Our chemical model has been recently extended with the option of a multilayer approach to ice chemistry \citep{Sipila16b}, but this issue is not considered here. Included desorption mechanisms are thermal desorption, cosmic-ray-induced desorption \citep{HH93}, and reactive desorption for exothermic surface reactions assuming $\sim$1\% efficiency \citep{Garrod07}. Quantum tunneling on grain surfaces is allowed for reactions with activation barriers, but diffusion is assumed to occur only thermally. Also, we do not consider grain coagulation, the inclusion of which would affect for example the charge balance in the gas phase \citep{Flower05}. A significant effect on the results presented here is not expected even if coagulation was included in the modeling, although a quantitative study would of course be needed to confirm this statement.

Both the gas-phase and grain-surface reaction networks used in this paper contain deuterated species with up to seven deuterium atoms, and explicit spin-state chemistry for $\rm H_2$, $\rm H_2^+$, $\rm H_3^+$, water and ammonia, and all of their deuterated isotopologs. Unlike in \citet{Sipila15a,Sipila15b}, we use in the present paper the KIDA gas-phase network \citep{Wakelam15} as the base upon which the deuterium and spin-state chemistry is added with the method described in full in \citet{Sipila15b}. The grain-surface network is essentially the same as in \citet{Sipila15b}, although it has been modifed slightly so that the (surface counterparts of) species not included in the KIDA network have been removed. The full gas-phase network consists of $\sim 77000$ reactions while the grain-surface network includes $\sim 2200$ reactions. However, in the present work we only consider subsets of the full network (see Sect.\,\ref{ss:models}). In the chemical calculations we assume that the gas is initially atomic with the exception of $\rm H_2$ and HD which are molecular. The adopted initial abundances are given in Table~\ref{tab1}.

\begin{table}
\caption{Initial chemical abundances with respect to $n_{\rm H}$, and the adopted initial $\rm H_2$ o/p ratio.}
\centering
\begin{tabular}{c c}
\hline \hline 
Species & Initial abundance  \\ \hline
$\rm H_2$ & 0.5 \\
$\rm He$ & $9.00\times10^{-2}$  \\
$\rm HD$ & $1.60\times10^{-5}$ \\
$\rm O$ & $2.56\times10^{-4}$ \\ 
$\rm C^+$ & $1.20\times10^{-4}$  \\
$\rm N$ & $7.60\times10^{-5}$ \\
$\rm S^+$ & $8.00\times10^{-8}$ \\
$\rm Si^+$ & $8.00\times10^{-9}$ \\
$\rm Na^+$ & $2.00\times10^{-9}$ \\
$\rm Mg^+$ & $7.00\times10^{-9}$ \\
$\rm Fe^+$ & $3.00\times10^{-9}$ \\
$\rm P^+$ & $2.00\times10^{-10}$ \\
$\rm Cl^+$ & $1.00\times10^{-9}$ \\
$\rm H_2\,(o/p)_{\rm ini}$ & $1.00\times10^{-1}$ \\
\hline
\end{tabular}
\label{tab1}
\end{table}

We follow the chemical evolution as a function of time and use the abundance gradients of selected cooling species as input to a Monte Carlo non-LTE radiative transfer program (Juvela, in prep.; see also \citealt{Juvela97}) to determine the total line cooling power ($\Lambda_{\rm line}$) at each time step. The included cooling species are presented in Sect.\,\ref{ss:models}. The gas is heated mainly by cosmic rays:
\begin{equation}
\Gamma_{\rm CR} = \Delta Q_{\rm CR} \, \zeta \, n({\rm H_2}) \, ,
\end{equation}
where $\Delta Q_{\rm CR} = 20 \, \rm eV$, $\zeta = 1.3 \times 10^{-17} \, \rm s^{-1}$ is the cosmic-ray ionization rate of $\rm H_2$ molecules, and $n({\rm H_2})$ is the $\rm H_2$ density \citep{Goldsmith78}\footnote{We note that \citet{Glassgold12} have calculated smaller values of $\Delta Q \sim 10 - 13 \, \rm eV$ for the medium densities probed here.}. Gas-dust collisional coupling may also heat or cool the gas at number densities above a few~$\times~10^4 \, \rm cm^{-3}$ \citep{Goldsmith01}:
\begin{align}
\Lambda_{\rm gd} &= 2.0 \times 10^{33} \, \left( n({\rm H_2}) / {\rm cm^{-3}} \right)^2 \, \left[(T_{\rm gas} - T_{\rm dust}) / {\rm K} \right] \nonumber \\
 &\times (T_{\rm gas} / 10 \, \rm K)^{0.5} \, erg \, cm^{-3} \, s^{-1} .
\end{align}
In addition, we include heating by the photoelectric effect ($\Gamma_{\rm peh}$; see \citealt{Juvela03b, Juvela11}). The photoelectric heating is calculated by another radiative transfer program \citep{Juvela05}. The unattenuated ISRF spectrum is taken from \citet{Black94}. We assume that the visual extinction at the edge of the model core is $A_{\rm V} = 2 \, \rm mag$. However, analogously to \citet{Sipila17}, we also added an extra layer outside the core corresponding to $A_{\rm V} = 1 \, \rm mag$ that is only used in the line cooling calculations. The abundances of the various species in this layer are assumed equal to the abundances at the core edge. If this extra layer is not included, the photon escape probability at the core edge is unphysically high because the radiative transfer program thinks there is nothing outside the model core to absorb the emitted photons, leading to low gas temperatures at the core edge. The extra layer is ignored in all other calculations besides line cooling.

Combining all of the above, the net cooling term in Eq.\,(\ref{eq_E}) becomes
\begin{equation}\label{heatcool}
\Lambda = \Gamma_{\rm CR} + \Gamma_{\rm peh} - \Lambda_{\rm gd} - \Lambda_{\rm line} \, .
\end{equation}
Finally we note that the dust temperature is here calculated with the radiative transfer program of \citet{Juvela05}, using dust opacity data from \citeauthor{OH94}~(\citeyear{OH94}; thin ice mantles). The dust temperature is determined before the chemical calculations, i.e., between steps (\ref{eq_mass}) and (\ref{eq_E}) in the calculation workflow.

\subsection{Chemical reaction sets}\label{ss:models}

\begin{table*}
\caption{Models discussed in this paper, along with descriptions of the chemical network and the cooling species included in each case.}
\centering
\begin{tabular}{c c c}
\hline \hline 
Network & Description & Cooling species included \\ \hline
A1 & Full gas-phase and grain-surface reaction networks described & $\rm ^{12}CO$, $\rm ^{13}CO$, $\rm C$, $\rm C^+$, $\rm O$, $\rm oH_2O$, p$\rm H_2O$, \\
	   & in Sect.\,\ref{ss:int_chem_rt}, but considering only species with up to five & $\rm NO$, $\rm HCN$, $\rm HNC$ \\
	   & atoms, $\sim$43500 (42000+1510) reactions in total \\ \hline
A1alt & As A1, but with some coolants removed & $\rm ^{12}CO$, $\rm ^{13}CO$, $\rm C$, $\rm C^+$, $\rm O$, $\rm oH_2O$, p$\rm H_2O$\\ \hline
B1 & Simple chemical network based on the one used by  & $\rm ^{12}CO$, $\rm ^{13}CO$, $\rm C$, $\rm C^+$, $\rm O$, $\rm H_2O^{(a)}$\\
    & \citeauthor{Nelson99} (\citeyear{Nelson99}; see Appendix~\ref{appendixa}), no\\
    & surface chemistry except for the formation of $\rm H_2$\\ \hline
B2 & As B1, but with more surface chemistry (see Appendix~\ref{appendixa}) & $\rm ^{12}CO$, $\rm ^{13}CO$, $\rm C$, $\rm C^+$, $\rm O$, $\rm H_2O^{(a)}$ \\ \hline
B2alt & As B2, but including photodesorption$^{(b)}$ of grain-surface CO and $\rm H_2O$ & $\rm ^{12}CO$, $\rm ^{13}CO$, $\rm C$, $\rm C^+$, $\rm O$, $\rm H_2O^{(a)}$ \\

\hline
\end{tabular}
\label{tab2}
\tablefoot{$^{(\rm a)}$ We assume that the $\rm H_2O$ o/p ratio is unity, and use the data available in LAMDA \citep{Schoier05} for $\rm oH_2O$ and $\rm pH_2O$. $^{(\rm b)}$ Photodesorption is not included in the other models. The assumed photodesorption yields are $2.7 \times 10^{-3}$ for CO \citep{Oberg09a} and $1.0 \times 10^{-3}$ for $\rm H_2O$ \citep{Oberg09b}.}
\end{table*}

In this paper we present the results of models that incorporate different chemical reaction sets and varying sets of cooling molecules. The models are tabulated in Table~\ref{tab2} along with a description of the network used in each case. The effect of the choice of cooling species is discussed in Sect.\,\ref{ss:dynamics}.

The main aim of the present paper is to study whether the chemical abundance gradients evolving dynamically with the core are significantly different from those deduced from static core models. To this end, we take three snapshots of the evolution of a collapsing core (see Sect.\,\ref{ss:collapsemodel}), and run a pseudo-time-dependent chemical model using the density and temperature structures corresponding to these snapshots, starting from the same initial chemical abundances as the dynamical model (Table~\ref{tab1}). In the static model, the physical quantities of the model do not evolve as the chemistry runs from $t = 0 \, \rm yr$ up to the snapshot time. The results of this analysis are presented in Sect.\,\ref{ss:staticmodels}.

\subsection{Efficiency of the code}\label{ss:codeeff}

The majority of the total running time of the code is spent on the chemical calculations. It takes $\sim$4 minutes to calculate the chemical evolution in one grid cell when using the gas-phase and grain-surface network~A1, depending also on the physical parameters (density, temperature etc.). The code is parallelized so that the chemical evolution can be computed simultaneously in multiple grid cells. The radiative transfer calculations, also parallelized, are completed in a timescale of two minutes per time step, so it is the chemistry that takes up the majority of the total computational time.

It takes about 14 hours of cpu time to calculate $1 \times 10^5 \, \rm yr$ of dynamical evolution using a standard desktop computer with four processor cores, using the A1 reaction set with (initial) time resolution of $\sim 10^3$\,yr (see Sect.\,\ref{ss:iso}). Therefore, models such as those presented in this paper can be run with a personal computer in reasonable timescale of a few days per model.

For this paper we did not perform calculations using the entire chemical reaction sets described in Sect.\,\ref{ss:int_chem_rt}. The advantage of using the full reaction scheme is that we could model the abundances of more complex species such as methanol, but at a great computational cost. Furthermore, the reaction set A1 used here captures all of the essential chemistry related to the cooling species (mainly CO, C, and water) and relatively simple species such as ammonia, and so our results, presented below, would remain unaffected even if we did switch to the full networks.

\section{Results}\label{s:results}

\begin{figure}
\centering
\includegraphics[width=1.0\columnwidth]{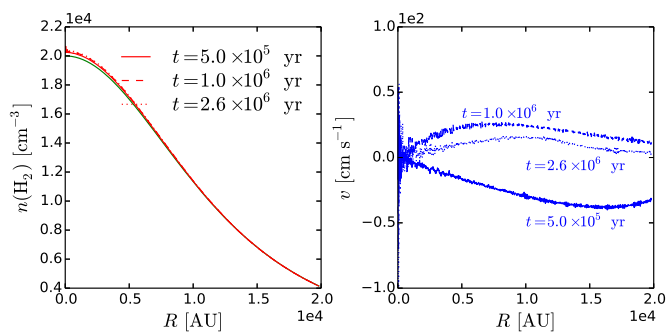}
\caption{The density ({\sl left panel}) and velocity profiles ({\sl right panel}) of a stable BES at different times, indicated in the figure. The green line shows the initial density profile. The velocity is initially zero across the core.
}
\label{fig:iso_stable}
\end{figure}

\begin{figure}
\centering
\includegraphics[width=1.0\columnwidth]{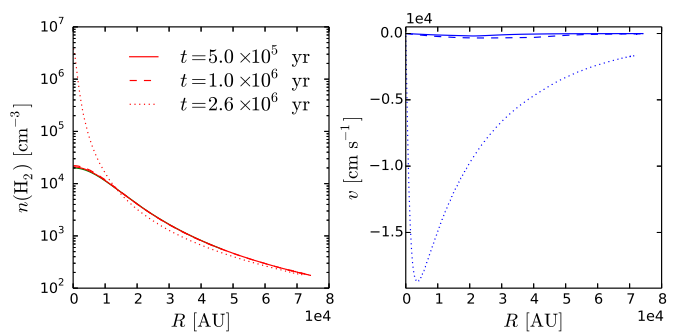}
\caption{The density ({\sl left panel}) and velocity profiles ({\sl right panel}) of an unstable BES at different times, indicated in the figure. The green line shows the initial density profile. The velocity is initially zero across the core.
}
\label{fig:iso_unstable}
\end{figure}

\subsection{Verification of the code: isothermal core}\label{ss:iso}

A basic requirement of the hydrodynamics code is that it should produce expected behavior for (isothermal) Bonnor-Ebert spheres: unstable cores should ultimately collapse while stable ones should not. To verify that the code works as intended, we carried out test calculations using two Bonnor-Ebert spheres with central density $n(\rm H_2) = 2\times10^4\,cm^{-3}$ and temperature $T = 10\,\rm K$. We picked two different values of the non-dimensional radius $\xi$ which represents the stability of the core: 4 (stable configuration) and 15 (unstable configuration); for more details on the stability of the Bonnor-Ebert sphere, see \citet{Bonnor56}. For the stable core, the $\rm H_2$ density at the edge is $n^{\rm edge}_{\rm H_2} \sim 4.2 \times 10^3 \, \rm cm^{-3}$, outer radius is $R_{\rm out} \sim19900 \, \rm AU$, and the mass is $M \sim 1.6 \, M_{\odot}$. For the unstable core, the corresponding parameters are $n^{\rm edge}_{\rm H_2} \sim 1.8 \times 10^2 \, \rm cm^{-3}$, $R_{\rm out} \sim74300 \, \rm AU$, and $M \sim 7.15 \, M_{\odot}$.

In Fig.\,\ref{fig:iso_stable} we show the evolution of the density and velocity profiles of the stable BES at different times during the dynamical evolution. As expected, the core does not collapse; instead, it oscillates. The velocity is initially zero across the core. At the very start of the dynamical evolution, a shallow infall profile (maximum infall speed of a few $\times$ $10\,\rm cm\,s^{-1}$) develops because of the gravitational potential (the core is not homogeneous), but this turns to an expansion in a timescale of $\sim 8 \times 10^5$\,yr. Later, periods of contraction and expansion alternate. The infall/expansion velocity stays below $\lesssim 1$ m\,s$^{-1}$ at all times. The plot also shows that the velocity profile fluctuates particularly at the inner boundary, but these fluctuations do not drive the evolution of the large-scale behavior. In this case the code runs until the termination time, $t = 5 \times 10^6\,\rm yr$, is reached.

Figure~\ref{fig:iso_unstable} shows three snapshots of the evolution of the unstable BES. Because the density contrast of this core is clearly higher than that of the stable BES (note the difference in density scale between Figs.~\ref{fig:iso_stable}~and~\ref{fig:iso_unstable}), a steeper infall profile develops early on and the core begins to collapse gradually. In this case the collapse is never halted because the core is gravitationally supercritical. The last time step shown in the figure corresponds to a time very close to when the code breaks the calculation loop (the flow becomes supersonic). Similar velocity fluctuations as those evident in Fig.\,\ref{fig:iso_stable} are present here as well, but they cannot be seen in the velocity scale of this figure. Overall, from the tests discussed above we conclude that the hydrodynamics part of the code works as intended.

The tests were performed adopting 1000 grid points for the spatial resolution, corresponding to an initial time step of $t \sim 1.1 \times 10^3$\,yr in the unstable core model. The choice of the time resolution affects the dynamics. If the initial time step is long, the core develops a steep infall profile quickly and collapses rapidly. The better the time resolution (i.e., the more grid points in the model), the more accurate the representation of the dynamics. However, using a small time step ($\gg$1000 grid points) increases the total calculational time tremendously when the chemistry is included. We studied the effect of the time resolution on the dynamical timescale by calculating the unstable BES model with different amounts of grid points. Table~\ref{tab3} summarizes the results of the test. The total running time of the model, defined by the time when the infalling flow becomes supersonic, increases with the number of grid points. The test results are close to each other (within a few tens of percent for the collapse timescale) when the number of grid points is of the order of 1000. Therefore we employ the value of 1000 also in the models presented below, as a compromise between accuracy of the dynamics and the required computational time. We stress that the collapse timescale itself is not the issue studied in this paper; we concentrate on the differences in chemical abundances between static and dynamical models, and this comparison is unaffected by how long it takes for the core to collapse.

\begin{table}
\caption{Summary of test runs exploring the effect of the spatial resolution on the collapse timescale.}
\centering
\begin{tabular}{c c c}
\hline \hline 
Grid points & $t_{\rm ini}$ & $t_{\rm f}$  \\ \hline
100 & $1.01\times10^4 \, \rm yr$ & $1.25\times10^6 \, \rm yr$\\
500 & $2.19\times10^3 \, \rm yr$ & $2.19\times10^6 \, \rm yr$\\
1000 & $1.01\times10^3 \, \rm yr$ & $2.61\times10^6 \, \rm yr$\\
3000 & $3.02\times10^2 \, \rm yr$ & $3.27\times10^6 \, \rm yr$\\
5000 & $2.01\times10^2 \, \rm yr$ & $3.54\times10^6 \, \rm yr$\\
\hline
\end{tabular}
\label{tab3}
\tablefoot{$t_{\rm ini}$ denotes the length of the initial time step; $t_{\rm f}$ denotes the final time, i.e., the time when the infall velocity becomes supersonic.}
\end{table}

\begin{figure*}
\centering
\includegraphics[width=2.0\columnwidth]{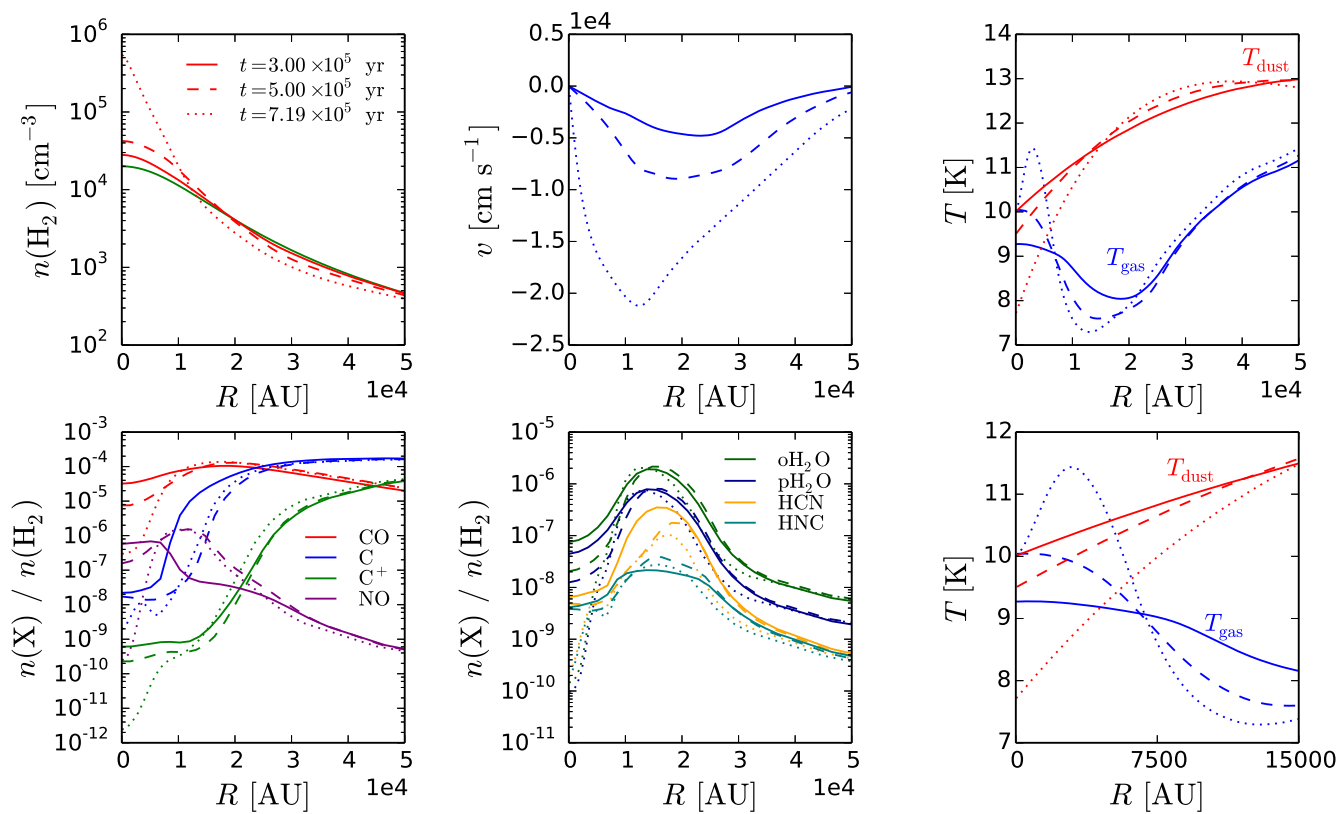}
\caption{{\sl Upper left:} density profiles of the collapsing model core at different times, indicated in the figure. The green line shows the initial density profile. {\sl Upper middle:} infall velocity profile of the model core at different times.  {\sl Upper right:} gas (blue) and dust (red) temperatures as functions of radius at different times. {\sl Lower left \& lower middle:} abundances of the cooling molecules as functions of radius at different times. {\sl Lower right:} gas and dust temperatures at different times, zoomed in to the innermost 15000\,AU of the core.
}
\label{fig:noniso}
\end{figure*}

\begin{figure*}
\centering
\includegraphics[width=1.8\columnwidth]{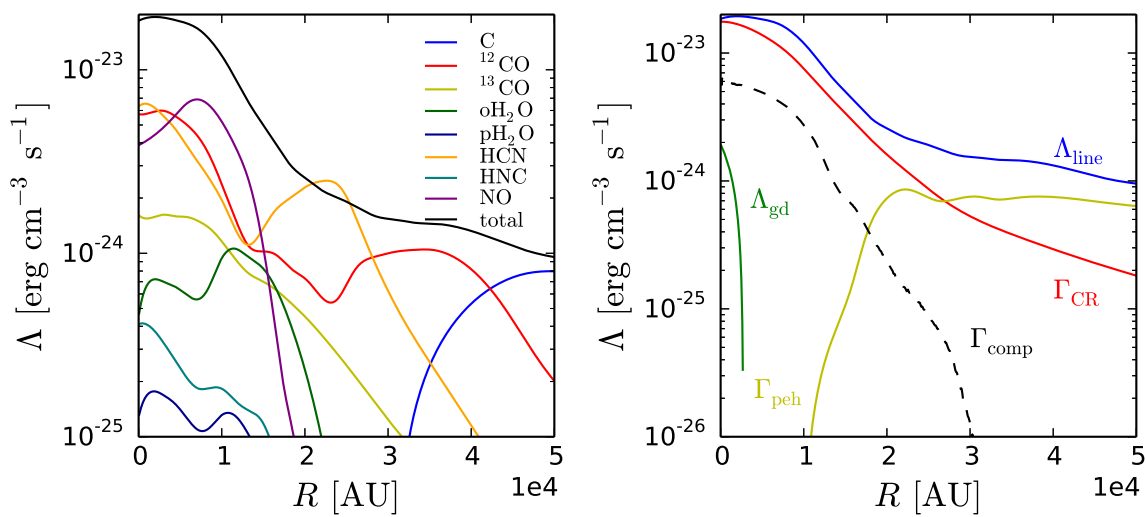}
\caption{{\sl Left:} cooling powers of the included cooling molecules as functions of radius. The total line cooling power ($\Lambda_{\rm line}$) is also shown in black. {\sl Right:} breakdown of the cooling and heating terms (indicated in the figure; see also Sect.\,\ref{ss:int_chem_rt}) as a function of radius. $\Gamma_{\rm comp}$ denotes the first term in Eq.\,(\ref{eq_E}), i.e. heating/cooling by gas compression/expansion, converted to the same units as the other heating and cooling terms (positive values indicate heating). The plots correspond to $t = 5 \times 10^5\,\rm yr$.
}
\label{fig:cooling}
\end{figure*}

\subsection{Collapse of a nonisothermal core}\label{ss:collapsemodel}

We present now the results of calculations from a model including chemistry and radiative transfer, i.e., in this case the core is no longer isothermal. We use the A1 model introduced in Sect.\,\ref{ss:models}. The initial core is the same as the unstable ($\xi = 15$) BES described in Sect.\,\ref{ss:iso}.

Figure~\ref{fig:noniso} shows three snapshots of the evolution in the innermost 50000\,AU of the model core, and the breakdowns of the net cooling power and line cooling power for one of these snapshot times. As was the case with the unstable BES discussed in Sect.\,\ref{ss:iso}, a minor infall motion was introduced already at the beginning of the simulation and the core started to collapse gradually. The infall flow became supersonic at $t = 7.19 \times 10^5\,\rm yr$, stopping the calculation.

The dust temperature is strongly tied to the density profile of the core. As the core becomes more centrally concentrated, the dust temperature drops at the center. The gas temperature is affected by several processes depending on the distance from the center of the core. Figure~\ref{fig:cooling} demonstrates the situation at $t = 5 \times 10^5 \, \rm yr$ so that one obtains an overall idea of the relative strengths of the heating and cooling processes. Looking first at the innermost 10000\,AU, it can be seen that the heating of the gas is dominated by cosmic rays and compression caused by the infall motion, while line radiation is mainly responsible for the cooling (the gas-dust coupling is also important near the origin). Outwards from 10000\,AU, the energy input from cosmic rays and compressive heating decrease more rapidly than the total line cooling power and, as a result, the gas is cooler than near the core center. Towards the core edge the gas temperature begins to rise again as photoelectric heating becomes important. We note that the gas and dust temperatures are not yet equal even at the core center when the code terminates. In the \citet{Goldsmith01} gas-dust coupling scheme, the two temperatures become coupled at a medium density of $\sim$$10^5 \, \rm cm^{-3}$ and finally equal to each other at a density of $\sim$$10^6 \, \rm cm^{-3}$ \citep{Goldsmith01, Keto05}. The central density in the A1 model at the last time step ($\sim$$5\times10^5 \, \rm cm^{-3}$) is below the equalization threshold.

From the breakdown of the line cooling power at this time step ($t = 5 \times 10^5 \, \rm yr$, corresponding to the dashed lines in Fig.\,\ref{fig:noniso}) one can see that the contributions of the cooling molecules to the total line cooling power are highly dependent on the distance from the core center. Cooling near the center of the core is dominated by HCN, CO, and NO, while atomic carbon is the most important coolant beyond $R \sim 40000$\,AU. ($\rm C^+$ is the most important coolant near the edge of the core, not shown.) Notably, NO and HCN are powerful coolants near the core center even though they are less abundant than CO by several orders of magnitude (Fig.\,\ref{fig:noniso}). For $t \lesssim 2 \times 10^5\, \rm yr$, before CO and particularly the late-type molecule NO have formed efficiently, the most important coolant in the central areas is HCN. Our results highlight the need for including a variety of cooling molecules when one employs a complex chemical model. In fact, it turns out that truncating the list of coolant species may increase the collapse timescale or prevent the collapse altogether. The effect of the line cooling scheme in our models is discussed further in Sect.\,\ref{ss:dynamics}. We note that the total line cooling power is lower than that of HCN alone at $R \sim 20000-25000$\,AU in Fig.\,\ref{fig:cooling}. This is because atomic carbon is providing heating (i.e., a negative contribution to cooling) in the region $R \sim 15000-30000$\,AU, caused by optical thickness effects: the line radiation is unable to escape the cloud, unlike in the outer regions where the medium density is low.

\subsection{Chemical abundances in the collapsing core}

\begin{figure*}
\centering
\includegraphics[width=2.0\columnwidth]{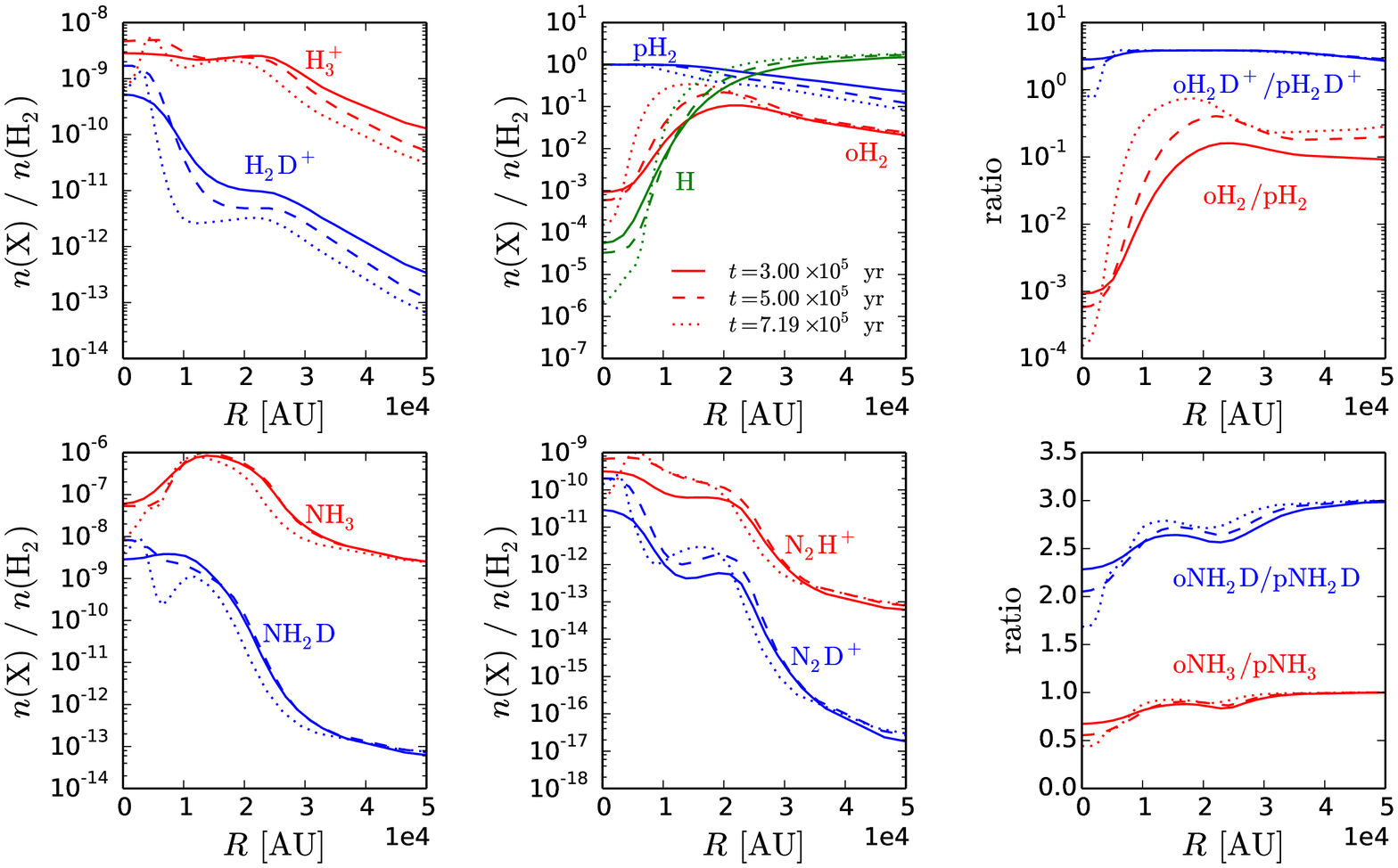}
\caption{Abundances (with respect to $\rm H_2$) and abundance ratios of various species in the collapsing model core at different times, indicated in the figure, as functions of radius. The abundances of the species represent sums over their respective spin states, when applicable, unless the spin state is explicitly displayed.
}
\label{fig:abus}
\end{figure*}

\begin{figure}
\centering
\includegraphics[width=0.9\columnwidth]{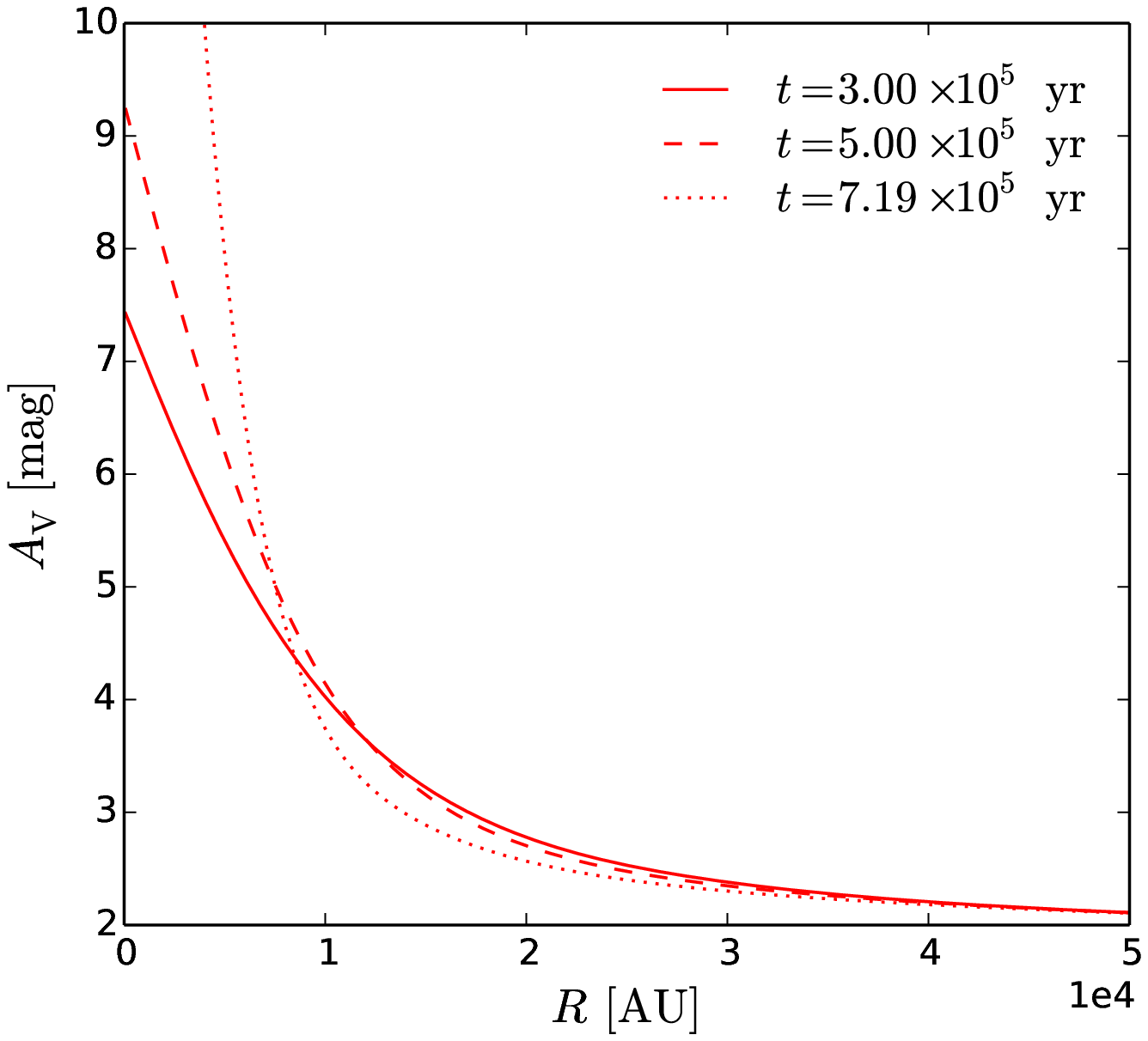}
\caption{Visual extinction $A_{\rm V}$ of the collapsing model core as a function of radius at different times, indicated in the figure.
}
\label{fig:Av}
\end{figure}

Figure~\ref{fig:abus} shows the abundances of several species of interest in the collapsing core at different times. A significant amount of ammonia is present in the gas from the center of the core to above 20000\,AU. The same is true for $\rm N_2H^+$ and $\rm H_3^+$. The abundances of the deuterated isotopologs of these species show a marked increase in abundance toward the center of the core. Just before the termination of the calculation, all of the aforementioned species are depleted near the center of the core due to the high medium density, either because of direct adsorption onto dust grains (neutrals) or secondary effects (ions).

Notably, the $\rm H_2$ ortho/para (hereafter o/p) ratio is $\sim$0.1 through most of the core, dropping to lower values only in the innermost 10000 or 20000\,AU depending on the time. The reason for this is the low visual extinction due to the low medium density. The visual extinction $A_{\rm V}$ is shown in Fig.\,\ref{fig:Av}. As the material flows inward, the density profile of the core becomes more centrally concentrated, and low values of $A_{\rm V}$ are found at increasingly smaller radii. At low $A_{\rm V}$, $\rm H_2$ is efficiently photodissociated and then reformed in the thermal o/p ratio of 3, while proton-exchange reactions of $\rm H_2$ with $\rm H^+$ and $\rm H_3^+$ drive the $\rm H_2$ o/p ratio towards the LTE value which is of the order of $10^{-3}$ at 20\,K. These competing processes lead to an $\rm H_2$ o/p ratio of $\sim 0.1$. Because of the close chemical relationship between the o/p ratio of $\rm H_2$ and that of $\rm H_2D^+$ \citep{Brunken14}, the $\rm H_2D^+$ o/p ratio stays near its thermal ratio of 3 as long as the $\rm H_2$ o/p ratio is high, dropping to lower values near the core center as the $\rm H_2$ o/p ratio decreases. We note that our code does not include $\rm H_2$ self-shielding. If this effect was included in the modeling, the $\rm H_2$ o/p ratio would decrease in the outer core along with the decreased $\rm H_2$ photodissociation rate. We tested the influence of $\rm H_2$ photodissociation on our results by running the A1 model with $\rm H_2$ photodissociation completely removed. The removal does not infuence the dynamics in any appreciable way. However, in this case the $\rm H_2$ o/p ratio is low ($\lesssim 10^{-3}$) throughout the core which can boost the deuteration fractions by more than two orders of magnitude in the outer core, depending on the species. This underlines the requirement of an accurate treatment of photodissociation when comparing models with observed abundance profiles (which is beyond the scope of this paper).

The o/p ratios of $\rm NH_3$ and $\rm NH_2D$ are determined by many factors \citep{Sipila15b}. In the outer core, the o/p ratios of both $\rm NH_3$ and $\rm NH_2D$ are thermal at all times. In the core center, the model predicts similar deviations from the thermal value as was found by \citet{Sipila15b} using static models. We compare predictions from our dynamical model and those of static models in detail in the next section. Abundance profiles for a selection of species other than those plotted above are shown and discussed in Sect.\,\ref{ss:abus}.

\subsection{Static models}\label{ss:staticmodels}

\begin{figure*}
\centering
\includegraphics[width=2.0\columnwidth]{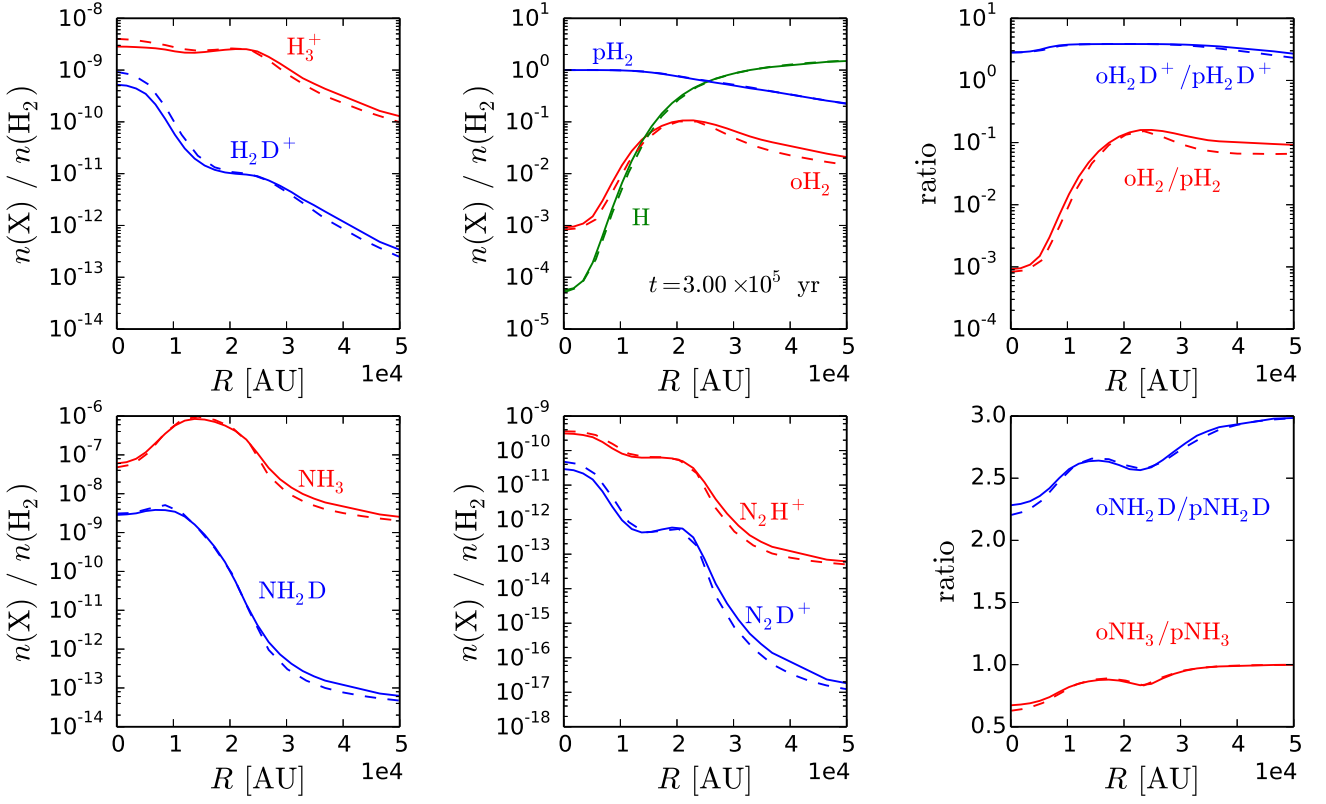}
\caption{Abundances (with respect to $\rm H_2$) and abundance ratios of various species in the collapsing model core (solid lines) and in a static core (dashed lines, see text) at $t = 3.00 \times 10^5 \, \rm yr$, as functions of radius. The abundances of the species represent sums over their respective spin states, when applicable, unless the spin state is explicitly displayed.
}
\label{fig:static1}
\end{figure*}

\begin{figure*}
\centering
\includegraphics[width=2.0\columnwidth]{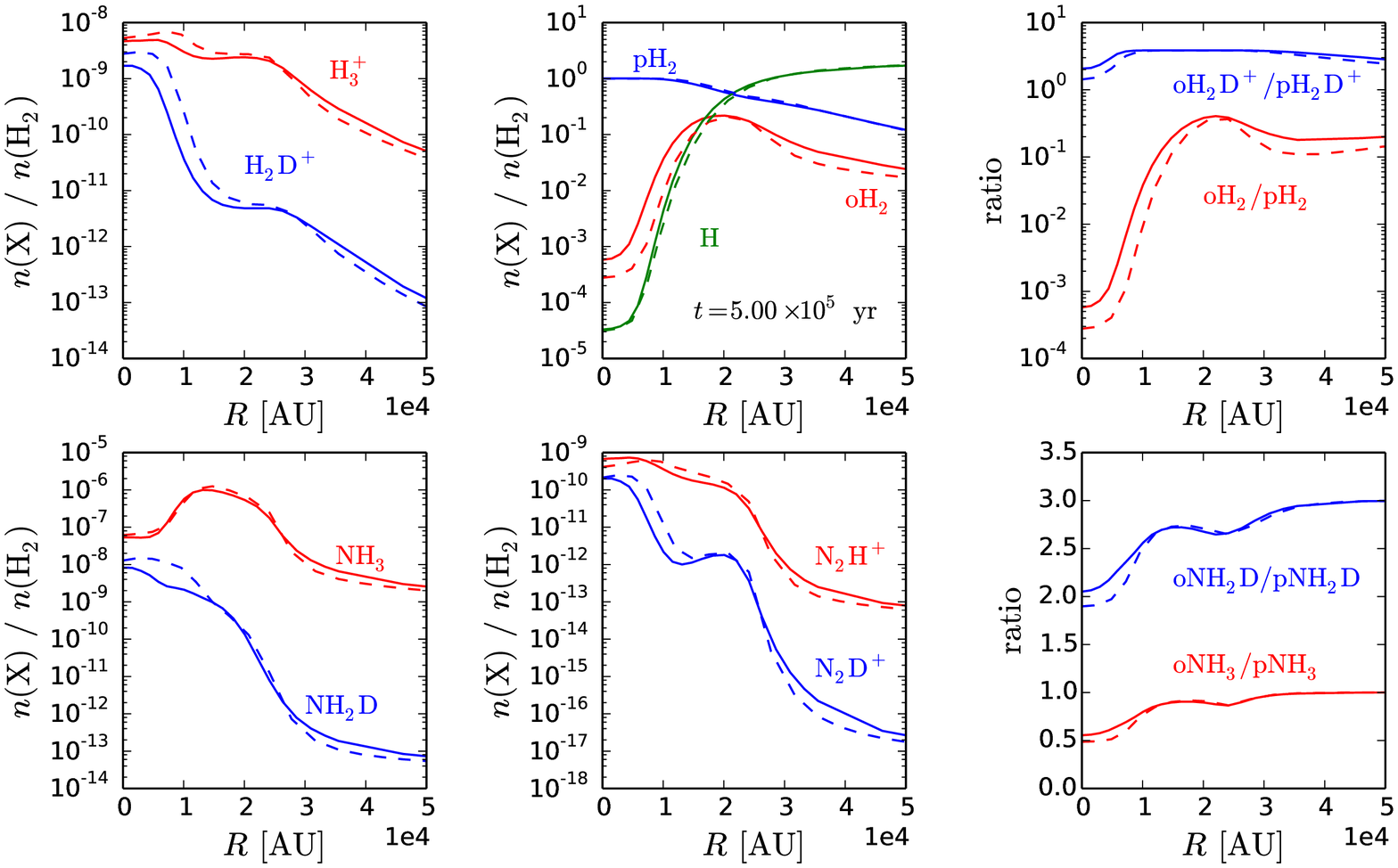}
\caption{Abundances (with respect to $\rm H_2$) and abundance ratios of various species in the collapsing model core (solid lines) and in a static core (dashed lines, see text) at $t = 5.00 \times 10^5 \, \rm yr$, as functions of radius. The abundances of the species represent sums over their respective spin states, when applicable, unless the spin state is explicitly displayed.
}
\label{fig:static2}
\end{figure*}

\begin{figure*}
\centering
\includegraphics[width=2.0\columnwidth]{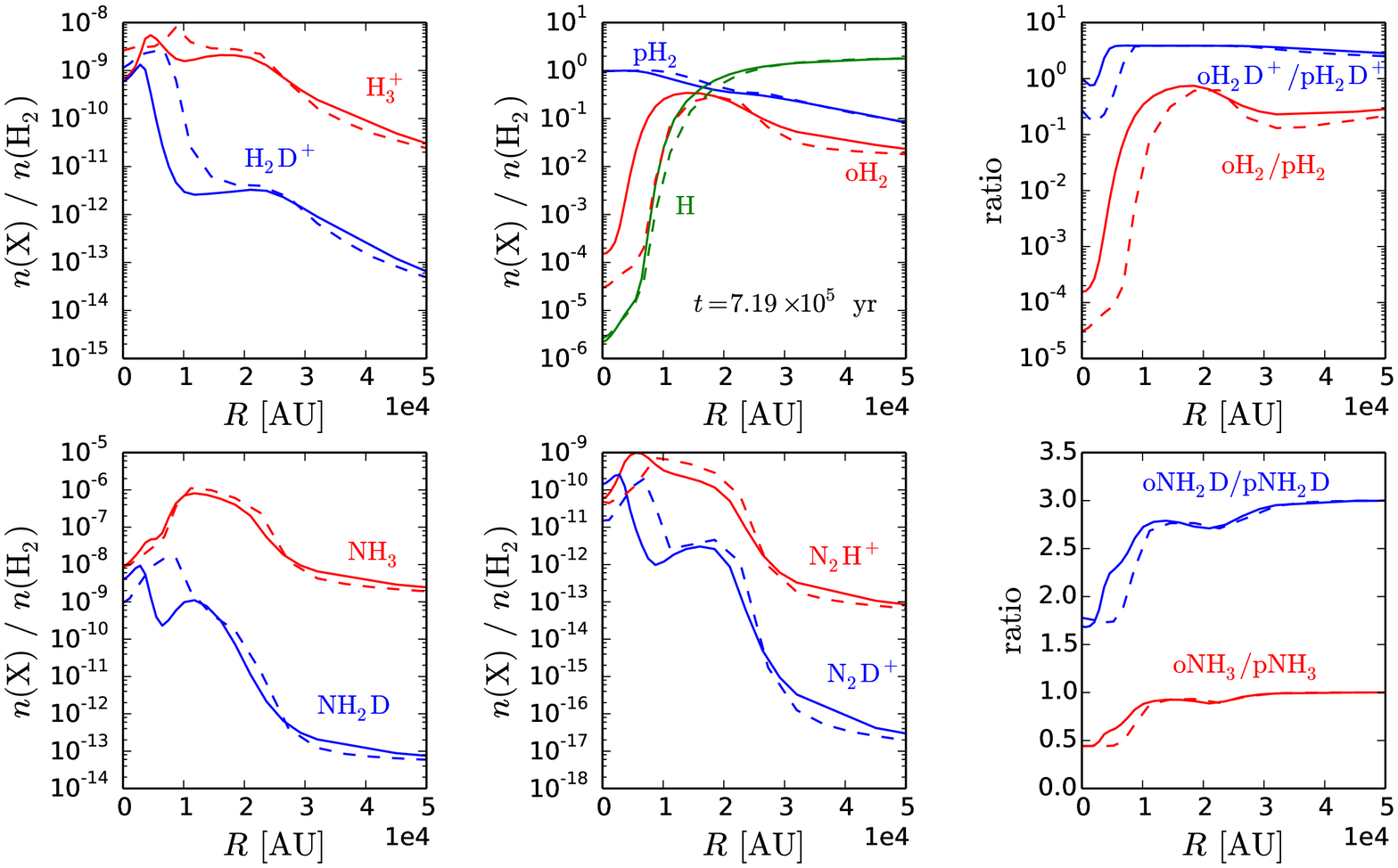}
\caption{Abundances (with respect to $\rm H_2$) and abundance ratios of various species in the collapsing model core (solid lines) and in a static core (dashed lines, see text) at $t = 7.18 \times 10^5 \, \rm yr$, as functions of radius. The abundances of the species represent sums over their respective spin states, when applicable, unless the spin state is explicitly displayed.
}
\label{fig:static3}
\end{figure*}

Figures~\ref{fig:static1}~to~\ref{fig:static3} show the abundances obtained from static models, calculated as described in Sect.\,\ref{ss:models}, compared to the abundances obtained from the dynamical model at different times. The results from the two types of model can be clearly different from each other, and the relative difference increases with time. At $t = 3.00 \times 10^5 \, \rm yr$ the results of the two models are nearly identical. However, at $t = 5.00 \times 10^5 \, \rm yr$ clear differences are already evident. The static model presents generally higher abundances and more extended profiles for most of the plotted species. We note that the high peak ammonia abundance (around $10^{-6}$) results from the adoption of the reactive desorption process, which leads to the desorption of the products of exothermic association reactions on grain surfaces with 1\% efficiency. If this desorption mechanism was turned off, the ammonia abundance would peak at lower values, but would still likely peak in the same location.

At $t = 7.19 \times 10^5 \, \rm yr$, the static model overproduces the abundances of many of the species with respect to the dynamical model outside the core center, and underproduces them near the origin. For example, the abundance of singly deuterated ammonia is an order of magnitude higher in the static model at $R \sim 10000$\,AU than in the dynamical model. The reason for the differences evident in the models is in the density and temperature profiles. In the static model, the core spends the entirety of its lifetime in a strongly centrally concentrated configuration which facilitates fast depletion onto grain surfaces in the central regions. On the other hand, the temperature near the center is higher than 10\,K throughout the chemical evolution of the core, as opposed to the dynamical model core where the temperature in the central areas is <10\,K for times up to $\lesssim 5 \times 10^5 \, \rm yr$. These effects are the cause of the variations evident in the abundance profiles in the two types of model.

The spin-state abundance ratios are also different in the two types of model at late times. The dynamical model does not yield an $\rm H_2D^+$ o/p ratio below $\sim$unity at any of the time steps displayed, whereas in the static model the $\rm H_2$ o/p ratio drops to such low values in the center that the $\rm H_2D^+$ o/p ratio decreases to $\sim$0.1. The discrepancy between the two models at $t = 7.19 \times 10^5 \, \rm yr$ is significant. The $\rm NH_3$ and $\rm NH_2D$ o/p ratios are also underestimated by the static model at late times, as compared to the dynamical model. These trends are opposite to what we found for the total abundances which are generally overestimated by the static model. The results from the dynamical model are especially interesting given that static models have trouble reproducing observed o/p ratios in (deuterated) ammonia \citep{Harju17a}, although we point out that we do not attempt to reproduce any particular observations in the present work. The differences in the $\rm H_2$ o/p ratio, and in the density and temperature profiles, lead to clear differences in the efficiency of deuteration in the two models. This point is further emphasized in Sect.\,\ref{ss:abus}.

In conclusion, our results show that the dynamical changes in temperature and density during the evolution of a molecular core are clearly reflected on the chemical abundance profiles, an effect which is completely missed by employing static models, and imply that the use of a static physical model is likely to lead to errors in the derivation of molecular abundances, for example. We discuss the observational impact of our modeling results in Sect.\,\ref{ss:obs}.

\section{Discussion}\label{s:discussion}

\subsection{Impact of chemistry on the dynamics}\label{ss:dynamics}

\begin{figure}
\centering
\includegraphics[width=1.0\columnwidth]{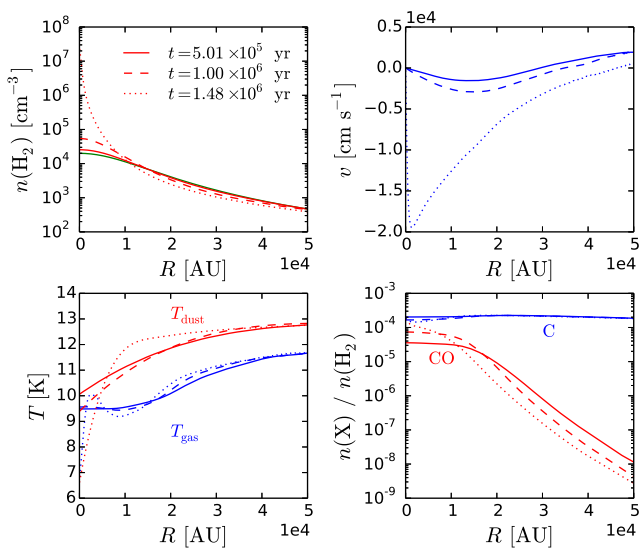}
\caption{As the top row and lower left panel of Fig.\,\ref{fig:noniso}, but calculated using reaction set B1.
}
\label{fig:noniso_simple}
\end{figure}

Calculating chemical development with an extensive chemical network is computationally very expensive, and so it is desirable to use a limited chemical network in 1D/3D hydrodynamical models that include a large amount of grid cells to keep the total computational time in a manageable scope. It is however not evident if and how the dynamics is altered if one switches from a simple reaction network, describing for example only $\rm H_2$ and CO formation, to a full gas-grain network. We quantify this issue next. We note that similar studies already exist in the literature: for example \citet{Hocuk14} and \citet{Hocuk16} have recently studied the effect of chemistry on the results of hydrodynamical models. However, even the complete network used in the latter work is very limited compared to that adopted in the present paper, justifying additional investigation into this issue.

We ran the collapsing core model (Sect.\,\ref{ss:collapsemodel}) using the simplified chemical networks~B1,~B2,~and~B2alt described in Sect.\,\ref{ss:models} and Appendix~\ref{appendixa}. The results of the calculation using the B1 reaction set are presented in Fig.\,\ref{fig:noniso_simple}. The collapse timescale of the core -- as determined by the termination condition of the code -- using reaction set B1 is $\sim$2.1 times that of the A1 model (Fig.\,\ref{fig:noniso}). The main reason for this difference is in the evolution of the C and CO abundances. In model~A1, CO forms (at early times) efficiently through neutral-neutral reactions such as $\rm O + C_2 \longrightarrow C + CO$ and $\rm O + CH_2 \longrightarrow CO + H + H$. These reactions are not included in model B1, where CO forms through $\rm HCO^+ + e^-$, and the CO abundance only becomes similar to that of C at very late times near the center of the core, and stays at very low abundances in the outer parts of the core throughout the time evolution. The cooling power of C is low near the core center ($\sim 2 \times 10^{-24}\,\rm erg \, cm^{-3} \, s^{-1}$) even though it is abundant, and so the slow increase in CO abundance means that it takes a significant amount of time for the total line cooling power to rise to high levels and for the collapse to proceed efficiently. The gas temperature profile behaves similarly to that in the A1 model (Fig.\,\ref{fig:noniso}), although the gas is overall warmer by about one K in the B1 model as long as the gas-dust coupling is not significant.

In model~B2 where surface reactions are included, oxygen is locked in grain-surface water in a relatively short timescale, resulting in low gas-phase abundances for CO and water. This translates to low line cooling rates and the core oscillates instead of collapsing. In the B2alt model where surface CO and water are also capable of photodesorbing, the gas cooling rates are higher and the core ultimately collapses. However, in this case the code terminates at $2.68 \times 10^6$\,yr, i.e., it takes a factor of $\sim1.8$ longer for the core to collapse than when using the B1 model scheme -- a factor of 3.8 higher than the collapse timescale of the A1 model.

We also ran the A1 model considering only the smaller set of coolants as in models B1 and B2 (marked as A1alt in Table~\ref{tab2}). In this case, the termination of the code is retarded to $2.93 \times 10^6$\,yr. A further test shows that if water is also removed from the set of coolants, the core never collapses. These results highlight the need to have a comprehensive description of cooling in order to obtain the most accurate estimate for the collapse timescale.

From the above it is clear that the dynamics of the core evolution can be strongly affected by the adopted chemical reaction network and by the set of cooling molecules used in the modeling. While disregarding depletion onto dust grains has an obvious impact on the modeling, it is also evident that limiting the gas-phase reaction network affects the results because of the differences in how the molecules are processed. A systematic study of the effect of considering chemical networks of varying complexity on the dynamics of core evolution, with the particular aim of obtaining information on the most essential processes, would be a worthwhile effort.

\subsection{Observational difference in dynamical and static models}\label{ss:obs}

\begin{figure*}
\centering
\includegraphics[width=2.0\columnwidth]{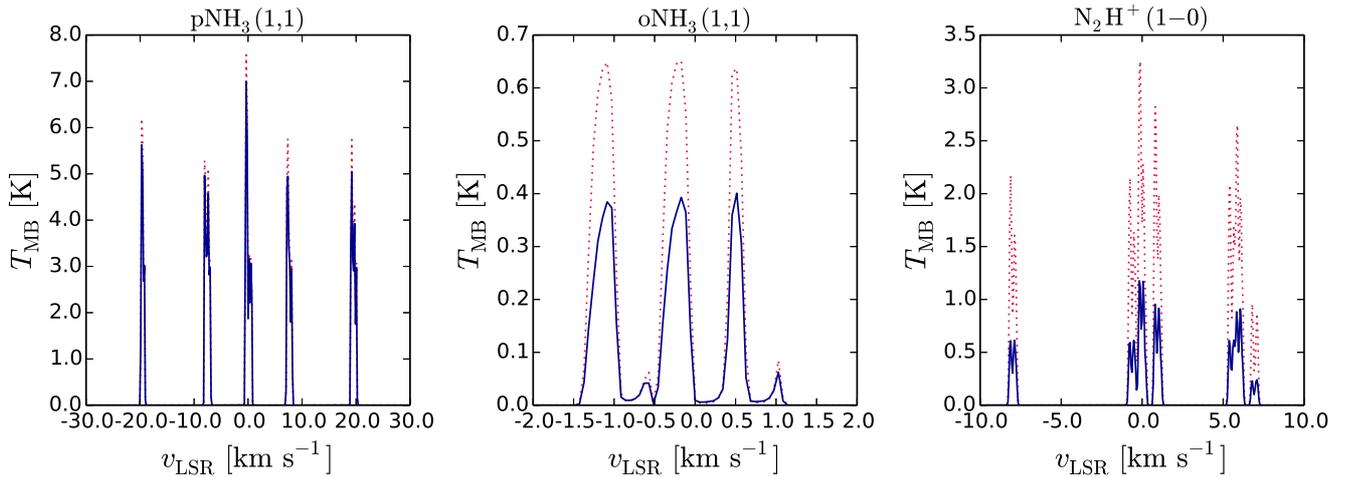}
\caption{Simulated line profiles of the $\rm pNH_3$(1,1), $\rm oNH_3$(1-0), and $\rm N_2H^+$($1-0$) transitions. The red dotted lines correspond to abundances obtained from the dynamical model, while the solid blue lines correspond to the static model. The evolutionary time in each case is $t = 7.19 \times 10^5 \, \rm yr$.
}
\label{fig:lines}
\end{figure*}

Figures~\ref{fig:static1} to \ref{fig:static3} exemplify that the abundance profiles obtained from dynamical models can differ clearly from those obtained from static models. The differences in the central areas are especially relevant for species with high critical densities, such as ammonia. In the previous figures we plotted the abundances of some common tracer species expected a priori to be important near the core center. To quantify the differences in line profiles resulting from different radial abundance profiles depending on the physical model, we calculated simulated line emission profiles for ammonia and $\rm N_2H^+$ using the radiative transfer program of \citet{Juvela97}. We modeled the (1,1) inversion transition of $\rm pNH_3$ at 23.69\,GHz, the ($1-0$) rotational transition of $\rm oNH_3$ at 572.50\,GHz, and the ($1-0$) rotational transition of $\rm N_2H^+$ at 93.17\,GHz. For simplicity, all of the simulated spectra were convolved to a 10$\arcsec$ beam. We assumed that the distance to the model core is 100\,pc.

Figure~\ref{fig:lines} shows the simulated line profiles at $t = 7.19 \times 10^5 \, \rm yr$. The difference between the static and dynamical models is striking. Although the static model predicts a higher abundance for all of the plotted species off the center of the core (Fig.\,\ref{fig:static3}), the emission is stronger in the dynamical model. This is because of the critical densities of the lines, which are $\sim$$2.0 \times 10^3 \, \rm cm^{-3}$, $\sim$$3.7 \times 10^7 \, \rm cm^{-3}$, and $\sim$$2.3 \times 10^5 \, \rm cm^{-3}$ (at 10\,K) for the modeled $\rm pNH_3$, $\rm oNH_3$, and $\rm N_2H^+$ transitions, respectively. For example, a density of $\sim 2.3 \times 10^5 \, \rm cm^{-3}$ corresponds to a radius of just a few thousand AU at this time step (Fig.\,\ref{fig:noniso}), meaning that the large part of the core where the $\rm N_2H^+$ abundance is higher in the static model is actually not emitting, leading to stronger emission in the dynamical model because of the higher abundance near the very center of the core. We note that the line simulations pertaining to static physical models adopt the infall velocity profile from the dynamical model at the appropriate time step, resulting in infall asymmetry in the static case as well.

These tests show that there can be a significant difference in simulated lines depending on whether one uses a dynamical or static physical model. Given the tendency of the static model to underpredict the various abundances near the core center and to overpredict them away from the center, the emission lines from the various high-density-tracing species will tend to be weak in the static model, while simultaneously the lower-density-tracing lines are likely to be too strong (optical depth effects will of course influence the situation as well). The requirement for an accurate model of the (thermal) history of a core is apparent.

Finally we comment on the recent paper by \citet{Caselli17}, who found that the abundance of $\rm oNH_3$ and the associated emission in the (1,1) line are not well fitted by a chemical model coupled with a static physical model of the structure of L1544 \citep{Keto10,Keto14}. In particular, the $\rm oNH_3$ abundance profile predicted by the model is at odds with the abundance profile deduced by \citet{Crapsi07}; the chemical model predicts strong ammonia depletion at the core center, and on the other hand too much ammonia toward the outer core. The results of the present paper imply that the situation could be remedied at least to some extent by employing the dynamical model, which predicts somewhat less depletion and also less ammonia in lower density gas. Quantitative conclusions on how the simulated lines would appear with the present model in the case of L1544 cannot be drawn without further modeling. We leave a detailed investigation of specific cores, such as L1544, for future work.

\subsection{Further abundance profiles in dynamical and static models}\label{ss:abus}

\begin{figure*}
\centering
\includegraphics[width=2.0\columnwidth]{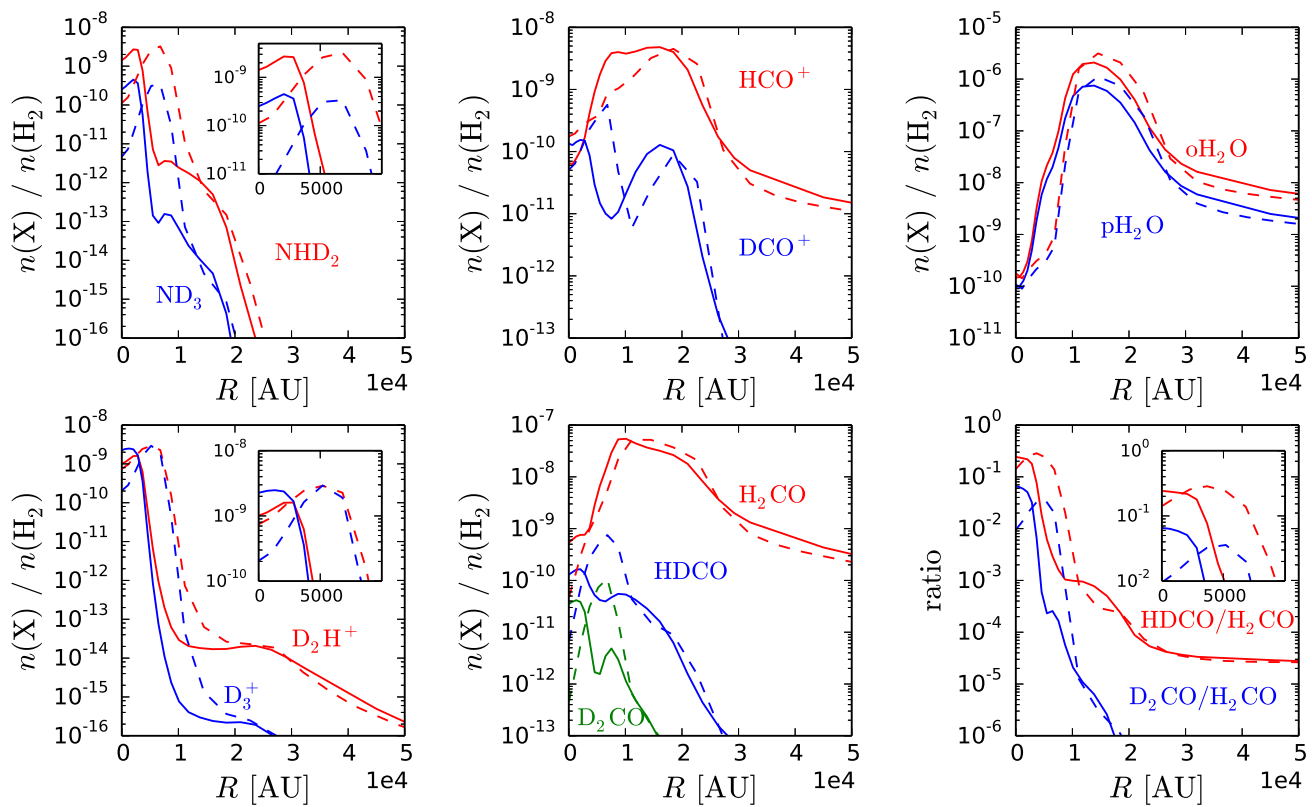}
\caption{As Fig.\,\ref{fig:static3}, but displaying additional species. Insets show zoomed-in views of the abundances in the innermost 10000~AU of the model core.
}
\label{fig:static_more}
\end{figure*}

In the preceding sections we showed the abundance profiles of selected species as given by the dynamical and static models at different time steps. For completeness, we show in Fig.\,\ref{fig:static_more} the abundance profiles of a variety of other species at $t = 7.19 \times 10^5 \, \rm yr$. The figure shows similar tendencies to Figs.\,\ref{fig:static1} to \ref{fig:static3}; the static model underpredicts abundances near the core center and overpredicts them in the outer core. This rule is not absolute, however, as evidenced by $\rm HCO^+$ which is more abundant in the central core in the static model than in the dynamical model, caused by a higher central gas-phase CO abundance in the static model (itself caused by the temperature which stays above 10\,K in the central areas throughout the chemical evolution in this particular static model, cf. the discussion in Sect.\,\ref{ss:staticmodels}).

Fig.\,\ref{fig:static_more} further emphasizes the fact that deuteration is overestimated in the central regions of the core by employing a static physical model. The overestimate increases with the number of deuterium atoms, suggesting potentially major implications for the interpretation of observations of multiply-deuterated species. Of course, the strength of the effect depends on the excitation conditions as well. This issue will be explored in detail in a future work. 

\section{Conclusions}\label{s:conclusions}

We presented a new one-dimensional spherically symmetric hydrodynamics code that integrates the solution of the basic hydrodynamics equations with radiative transfer calculations and a gas-grain chemical code that includes deuterium and spin-state chemistry. Time-dependent abundances from the chemical model are used as input to the radiative transfer calculations so that the cooling rates can be determined self-consistently. Our main aim in the present paper was to quantify whether the molecular abundances predicted by a model including dynamical evolution differ from those obtained with static physical models. The full gas-phase and grain-surface reaction sets considered here contain a combined total of $\sim$43500 reactions. The use of an extensive description of chemistry allows investigation of the effect of chemistry on the overall dynamics, when our fiducial model results are compared against those obtained with alternative chemical reaction sets with decreased degrees of complexity. We followed the dynamical evolution up to the point when the infall velocity becomes supersonic. The formation of a central source is not presently included in the code.

We found that the difference in predicted abundances between the dynamical and static models is generally significant, but this result is time-dependent. At early times, well before the core starts to collapse, the results from the two models are nearly identical. As the infall velocity increases and the core contracts, the modeling results begin to diverge. At late times the static models generally overpredict abundances in the outer parts of the core, and underpredict them near the core center, with respect to the dynamical model. The differences in the abundances can translate to striking differences in simulated emission lines. For the ammonia and $\rm N_2H^+$ lines simulated here, the dynamical model predicts clearly stronger emission than the static model because the lines originate near the center of the core where the dynamical model predicts higher abundances. Low-density-tracing lines are accordingly expected to appear clearly brighter in the static model.

Our results show that using static physical models on the one hand to simulate abundance gradients and on the other hand to reproduce observed line emission profiles may lead to large errors if the core has already started contracting. Also, an extensive chemical network is needed for the most accurate representation of the dynamics because of the effect of the chemistry on the total line cooling power; the collapse timescale depends on the number of grid points in the model. The downside of the present approach where the chemistry and line cooling are calculated self-consistently is that the total calculation time becomes very long if the time step is small. However, we found that the core collapse timescale reaches a plateau with increasing number of grid points, so that a fairly good approximation of the dynamics can be reached with a limited number of grid points, keeping the total calculation time in a manageable scope.

We also studied the effect of the complexity of the adopted chemical networks on the overall dynamical evolution. We found that using a very simple gas-phase chemical network limited to essential CO and water chemistry increases the collapse timescale of our model core by a factor of $\sim$2.1 with respect to the model with complete gas-grain chemistry. On other hand, adding a very simple set of grain-surface reactions to the gas-phase network leads to a solution where the core does not collapse, caused by the trapping of oxygen into grain-surface water, unless photodesorption of water (and CO) is included. In the latter case the collapse timescale is $\sim$3.7 times that of the model with full chemistry, indicating that the use of very simple chemical networks in hydrodynamical simulations may lead to overestimation of the collapse timescale.

Finally, we found that in a collapse model including complex chemistry, it is vital to consider a variety of cooling species. In the central regions, most of the molecular line cooling is in our model due to HCN, CO, and NO. If only a limited set of coolants is considered, the core collapse timescale increases or the collapse is prevented altogether because of insufficient cooling in the central regions of the core.

Our hydrodynamical scheme is based on thermal processes and does not at present include additional sources of pressure such as the magnetic field or turbulence. With the addition of these processes, left for future work, the versatility of our model would be improved as we would obtain a more realistic description of the physics of the collapse. This will be beneficial when real cores are simulated. However, the main conclusions of the present paper on the discrepancy between the dynamical and static models would not be affected. The code can also be extended to model the collapse up to the formation of a central source. This would yield interesting information on chemical abundances in the innermost areas of collapsing cores, on the sub-100\,AU scale.

\begin{acknowledgements}
We thank the anonymous referee for comments that helped to improve the paper. We also thank Mika Juvela for valuable discussions on the implementation of radiative transfer in the modeling. P.C. acknowledges financial support of the European Research Council (ERC; project PALs 320620).
\end{acknowledgements}

\bibliographystyle{aa}
\bibliography{hydro.bib}

\appendix

\onecolumn

\section{Chemical networks B1 and B2}\label{appendixa}

In this Appendix, we present the gas-phase and grain-surface networks that are used in the B1, B2, and B2alt models summarized in Table~\ref{tab2}. The gas-phase network given in Table~\ref{tabb1} is used in all of the B models. The difference between the B1 and B2 cases is in the surface chemistry; in model B1 only surface formation of $\rm H_2$ is considered, while models B2 and B2alt include all of the surface reactions given in Table~\ref{tabb2}.

\begin{longtable}{ccccccccccc}
\caption{Gas-phase reaction network used in models B1, B2, and B2alt. The form of the rate coefficient is given by the column labeled "type": (1) $k~=~\alpha\zeta$, where $\zeta$ is the cosmic ray ionization rate; (2) $k = \alpha \exp(-\gamma A_{\rm V})$, where $A_{\rm V}$ is the visual extinction; (3) $k = \alpha (T/300)^\beta \exp(-\gamma/T)$; (4) $k = \alpha \beta (0.62 + 0.4767\gamma (300/T)^{1/2})$; (5)~$k = \alpha\beta(1 + 0.0967\gamma (300/T)^{1/2} + (\gamma^2/10.526) (300/T))$.}\\
\hline\hline
\centering

& & Chemical reaction & & & & $\alpha$ & $\beta$ & $\gamma$ & Type \\ \hline
\endfirsthead
\caption{continued.}\\
\hline\hline
& & Chemical reaction & & & & $\alpha$ & $\beta$ & $\gamma$ & Type \\ \hline
\endhead
$\rm H     $ & $\rm CRP   $ & $\longrightarrow$  & $\rm H^+   $ & $\rm e^-   $ & & 4.60e-01 & 0.00e+00 & 0.00e+00 & $1$\\ 
$\rm He    $ & $\rm CRP   $ & $\longrightarrow$  & $\rm He^+  $ & $\rm e^-   $ & & 5.00e-01 & 0.00e+00 & 0.00e+00 & $1$\\ 
$\rm H_2   $ & $\rm CRP   $ & $\longrightarrow$  & $\rm H     $ & $\rm H     $ & & 1.00e-01 & 0.00e+00 & 0.00e+00 & $1$\\ 
$\rm H_2   $ & $\rm CRP   $ & $\longrightarrow$  & $\rm H     $ & $\rm H^+   $ & $\rm e^-   $ & 2.20e-02 & 0.00e+00 & 0.00e+00 & $1$ \\ 
$\rm H_2   $ & $\rm CRP   $ & $\longrightarrow$  & $\rm H_2^+ $ & $\rm e^-   $ & & 9.30e-01 & 0.00e+00 & 0.00e+00 & $1$\\ 
$\rm C     $ & $\rm CRP   $ & $\longrightarrow$  & $\rm C^+   $ & $\rm e^-   $ & & 1.02e+03 & 0.00e+00 & 0.00e+00 & $1$\\ 
$\rm OH    $ & $\rm CRP   $ & $\longrightarrow$  & $\rm H     $ & $\rm O     $ & & 5.10e+02 & 0.00e+00 & 0.00e+00 & $1$\\ 
$\rm H_2O  $ & $\rm CRP   $ & $\longrightarrow$  & $\rm H     $ & $\rm OH    $ & & 9.70e+02 & 0.00e+00 & 0.00e+00 & $1$\\ 
$\rm CO    $ & $\rm CRP   $ & $\longrightarrow$  & $\rm C     $ & $\rm O     $ & & 5.00e+00 & 0.00e+00 & 0.00e+00 & $1$\\ 
$\rm CO    $ & $\rm CRP   $ & $\longrightarrow$  & $\rm CO^+  $ & $\rm e^-   $ & & 3.00e+00 & 0.00e+00 & 0.00e+00 & $1$\\ 
$\rm H_2   $ & $\rm PHOTON$ & $\longrightarrow$  & $\rm H     $ & $\rm H     $ & & 3.40e-11 & 0.00e+00 & 2.50e+00 & $2$\\ 
$\rm C     $ & $\rm PHOTON$ & $\longrightarrow$  & $\rm C^+   $ & $\rm e^-   $ & & 3.10e-10 & 0.00e+00 & 3.33e+00 & $2$\\ 
$\rm OH    $ & $\rm PHOTON$ & $\longrightarrow$  & $\rm H     $ & $\rm O     $ & & 3.90e-10 & 0.00e+00 & 2.24e+00 & $2$\\ 
$\rm CO    $ & $\rm PHOTON$ & $\longrightarrow$  & $\rm C     $ & $\rm O     $ & & 2.60e-10 & 0.00e+00 & 3.53e+00 & $2$\\ 
$\rm HCO^+ $ & $\rm PHOTON$ & $\longrightarrow$  & $\rm H^+   $ & $\rm CO    $ & & 1.50e-10 & 0.00e+00 & 2.50e+00 & $2$\\ 
$\rm Fe    $ & $\rm PHOTON$ & $\longrightarrow$  & $\rm Fe^+  $ & $\rm e^-   $ & & 2.80e-10 & 0.00e+00 & 2.20e+00 & $2$\\ 
$\rm H_2O  $ & $\rm PHOTON$ & $\longrightarrow$  & $\rm H     $ & $\rm OH    $ & & 8.00e-10 & 0.00e+00 & 2.20e+00 & $2$\\ 
$\rm H_2O  $ & $\rm PHOTON$ & $\longrightarrow$  & $\rm H_2O^+$ & $\rm e^-   $ & & 3.10e-11 & 0.00e+00 & 3.90e+00 & $2$\\ 
$\rm H^+   $ & $\rm e^-   $ & $\longrightarrow$  & $\rm H     $ & $\rm PHOTON$ & & 3.50e-12 & -7.00e-01 & 0.00e+00 & $3$\\ 
$\rm H_2^+ $ & $\rm e^-   $ & $\longrightarrow$  & $\rm H     $ & $\rm H     $ & & 2.53e-07 & -5.00e-01 & 0.00e+00 & $3$\\ 
$\rm H_3^+ $ & $\rm e^-   $ & $\longrightarrow$  & $\rm H     $ & $\rm H     $ & $\rm H     $ & 4.36e-08 & -5.20e-01 & 0.00e+00 & $3$ \\ 
$\rm H_3^+ $ & $\rm e^-   $ & $\longrightarrow$  & $\rm H     $ & $\rm H_2   $ & & 2.34e-08 & -5.20e-01 & 0.00e+00 & $3$\\ 
$\rm He^+  $ & $\rm e^-   $ & $\longrightarrow$  & $\rm He    $ & $\rm PHOTON$ & & 4.50e-12 & -6.70e-01 & 0.00e+00 & $3$\\ 
$\rm HCO^+ $ & $\rm e^-   $ & $\longrightarrow$  & $\rm H     $ & $\rm CO    $ & & 2.80e-07 & -6.90e-01 & 0.00e+00 & $3$\\ 
$\rm C^+   $ & $\rm e^-   $ & $\longrightarrow$  & $\rm C     $ & $\rm PHOTON$ & & 4.40e-12 & -6.10e-01 & 0.00e+00 & $3$\\ 
$\rm Fe^+  $ & $\rm e^-   $ & $\longrightarrow$  & $\rm Fe    $ & $\rm PHOTON$ & & 3.70e-12 & -6.50e-01 & 0.00e+00 & $3$\\ 
$\rm CH^+  $ & $\rm e^-   $ & $\longrightarrow$  & $\rm C     $ & $\rm H     $ & & 7.00e-08 & -5.00e-01 & 0.00e+00 & $3$\\ 
$\rm OH^+  $ & $\rm e^-   $ & $\longrightarrow$  & $\rm H     $ & $\rm O     $ & & 6.30e-09 & -4.80e-01 & 0.00e+00 & $3$\\ 
$\rm CO^+  $ & $\rm e^-   $ & $\longrightarrow$  & $\rm C     $ & $\rm O     $ & & 2.75e-07 & -5.50e-01 & 0.00e+00 & $3$\\ 
$\rm H_2O^+$ & $\rm e^-   $ & $\longrightarrow$  & $\rm O     $ & $\rm H_2   $ & & 3.90e-08 & -5.00e-01 & 0.00e+00 & $3$\\ 
$\rm H_2O^+$ & $\rm e^-   $ & $\longrightarrow$  & $\rm H     $ & $\rm OH    $ & & 8.60e-08 & -5.00e-01 & 0.00e+00 & $3$\\ 
$\rm H_2O^+$ & $\rm e^-   $ & $\longrightarrow$  & $\rm H     $ & $\rm H     $ & $\rm O     $ & 3.05e-07 & -5.00e-01 & 0.00e+00 & $3$ \\ 
$\rm H_3O^+$ & $\rm e^-   $ & $\longrightarrow$  & $\rm H     $ & $\rm H     $ & $\rm OH    $ & 2.60e-07 & -5.00e-01 & 0.00e+00 & $3$ \\ 
$\rm H_3O^+$ & $\rm e^-   $ & $\longrightarrow$  & $\rm H     $ & $\rm H_2O  $ & & 1.10e-07 & -5.00e-01 & 0.00e+00 & $3$\\ 
$\rm H_3O^+$ & $\rm e^-   $ & $\longrightarrow$  & $\rm H_2   $ & $\rm OH    $ & & 6.00e-08 & -5.00e-01 & 0.00e+00 & $3$\\ 
$\rm H_3O^+$ & $\rm e^-   $ & $\longrightarrow$  & $\rm H     $ & $\rm O     $ & $\rm H_2   $ & 5.60e-09 & -5.00e-01 & 0.00e+00 & $3$ \\ 
$\rm H_2   $ & $\rm H_2^+ $ & $\longrightarrow$  & $\rm H     $ & $\rm H_3^+ $ & & 2.10e-09 & 0.00e+00 & 0.00e+00 & $3$\\ 
$\rm H     $ & $\rm H_2^+ $ & $\longrightarrow$  & $\rm H_2   $ & $\rm H^+   $ & & 6.40e-10 & 0.00e+00 & 0.00e+00 & $3$\\ 
$\rm C     $ & $\rm H_3^+ $ & $\longrightarrow$  & $\rm H_2   $ & $\rm CH^+  $ & & 2.00e-09 & 0.00e+00 & 0.00e+00 & $3$\\ 
$\rm O     $ & $\rm H_3^+ $ & $\longrightarrow$  & $\rm H_2   $ & $\rm OH^+  $ & & 7.98e-10 & -1.56e-01 & 1.41e+00 & $3$\\ 
$\rm CO    $ & $\rm H_3^+ $ & $\longrightarrow$  & $\rm H_2   $ & $\rm HCO^+ $ & & 9.45e-01 & 1.99e-09 & 2.51e-01 & $5$\\ 
$\rm Fe    $ & $\rm H_3^+ $ & $\longrightarrow$  & $\rm H     $ & $\rm H_2   $ & $\rm Fe^+  $ & 4.90e-09 & 0.00e+00 & 0.00e+00 & $3$ \\ 
$\rm H_2   $ & $\rm He^+  $ & $\longrightarrow$  & $\rm H     $ & $\rm He    $ & $\rm H^+   $ & 3.30e-15 & 0.00e+00 & 0.00e+00 & $3$ \\ 
$\rm CO    $ & $\rm He^+  $ & $\longrightarrow$  & $\rm He    $ & $\rm O     $ & $\rm C^+   $ & 1.00e+00 & 1.75e-09 & 2.51e-01 & $5$ \\ 
$\rm H_2   $ & $\rm He^+  $ & $\longrightarrow$  & $\rm He    $ & $\rm H_2^+ $ & & 9.60e-15 & 0.00e+00 & 0.00e+00 & $3$\\ 
$\rm H_2   $ & $\rm C^+   $ & $\longrightarrow$  & $\rm H     $ & $\rm CH^+  $ & & 1.50e-10 & 0.00e+00 & 4.64e+03 & $3$\\ 
$\rm OH    $ & $\rm C^+   $ & $\longrightarrow$  & $\rm H     $ & $\rm CO^+  $ & & 1.00e+00 & 9.15e-10 & 5.50e+00 & $4$\\ 
$\rm H_2   $ & $\rm CO^+  $ & $\longrightarrow$  & $\rm H     $ & $\rm HCO^+ $ & & 7.50e-10 & 0.00e+00 & 0.00e+00 & $3$\\ 
$\rm H_2   $ & $\rm OH^+  $ & $\longrightarrow$  & $\rm H     $ & $\rm H_2O^+$ & & 1.10e-09 & 0.00e+00 & 0.00e+00 & $3$\\ 
$\rm H_2   $ & $\rm H_2O^+$ & $\longrightarrow$  & $\rm H     $ & $\rm H_3O^+$ & & 6.10e-10 & 0.00e+00 & 0.00e+00 & $3$\\ \hline\\
$\rm H_2O  $ & $\rm H_3^+ $ & $\longrightarrow$  & $\rm H_2   $ & $\rm H_3O^+$ & & 1.00e+00 & 1.73e-09 & 5.41e+00 & $4$\\ 
$\rm C     $ & $\rm OH    $ & $\longrightarrow$  & $\rm H     $ & $\rm CO    $ & & 1.15e-10 & -3.39e-01 & -1.08e-01 & $3$\\ 
$\rm H     $ & $\rm CO^+  $ & $\longrightarrow$  & $\rm CO    $ & $\rm H^+   $ & & 4.00e-10 & 0.00e+00 & 0.00e+00 & $3$\\ 
\hline

\label{tabb1}
\end{longtable}

\begin{longtable}{ccccccc}
\caption{Grain-surface network used in models B2 and B2alt. The rate coefficients for the surface reactions are constructed as discussed in \citet{Sipila15a}; a detailed presentation is omitted here for brevity. For the cosmic-ray dissociation and photodissociation reactions, the meaning of the $\alpha$ and $\beta$ coefficients is the same as for the corresponding gas-phase reactions (Table~\ref{tabb1}). For the other reactions $\alpha$ and $\beta$ represent the branching ratio and activation energy, respectively.}\\
\hline\hline
\centering

& & Chemical reaction & & &  $\alpha$ & $\beta$ \\ \hline
\endfirsthead
\caption{continued.}\\
\hline\hline
& & Chemical reaction & & &  $\alpha$ & $\beta$ \\ \hline
\endhead
$\rm CO*   $ & $\rm CRP   $ & $\longrightarrow$  & $\rm C*    $ & $\rm O*    $ & 5.00e+00 & 0.00e+00 \\ 
$\rm OH*   $ & $\rm CRP   $ & $\longrightarrow$  & $\rm O*    $ & $\rm H*    $ & 5.10e+02 & 0.00e+00 \\ 
$\rm H_2O* $ & $\rm CRP   $ & $\longrightarrow$  & $\rm OH*   $ & $\rm H*    $ & 9.70e+02 & 0.00e+00 \\ 
$\rm CO*   $ & $\rm PHOTON$ & $\longrightarrow$  & $\rm C*    $ & $\rm O*    $ & 2.60e-10 & 3.53e+00 \\ 
$\rm OH*   $ & $\rm PHOTON$ & $\longrightarrow$  & $\rm O*    $ & $\rm H*    $ & 3.90e-10 & 2.24e+00 \\ 
$\rm H_2O* $ & $\rm PHOTON$ & $\longrightarrow$  & $\rm OH*   $ & $\rm H*    $ & 8.00e-10 & 2.20e+00 \\ 
$\rm C*    $ & $\rm O*    $ & $\longrightarrow$  & $\rm CO*   $ & & 1.00e+00 & 0.00e+00 \\ 
$\rm H*    $ & $\rm H*    $ & $\longrightarrow$  & $\rm H_2*  $ & & 1.00e+00 & 0.00e+00 \\ 
$\rm H*    $ & $\rm O*    $ & $\longrightarrow$  & $\rm OH*   $ & & 1.00e+00 & 0.00e+00 \\ 
$\rm H*    $ & $\rm OH*   $ & $\longrightarrow$  & $\rm H_2O* $ & & 1.00e+00 & 0.00e+00 \\ 
$\rm C*    $ & $\rm OH*   $ & $\longrightarrow$  & $\rm CO*   $ & $\rm H*    $ & 1.00e+00 & 0.00e+00 \\ 
\hline

\label{tabb2}
\end{longtable}

\end{document}